# Calypso Venus Scout

A white paper submitted to

The Committee on the Planetary Science Decadal Survey (2023-2032) of

The National Academies of Sciences

Philip Horzempa

LeMoyne College; Syracuse, New York

August 15, 2020

email: horzempa45@gmail.com

Telephone:  1-3157951259

Acknowledgement: Sam Zaref for his depiction of the Calypso Venus Scout

# CALYPSO VENUS SCOUT

# MISSION PROPOSAL for 2020 PLANETARY DECADAL SURVEY

The Calypso Venus Scout is a mobile, low-altitude survey and mapping mission. A unique design allows the science payload to view a significant amount of the surface of Venus from an altitude of 10-25 km. The harsh environment of the planet makes a surface rover or a low-altitude balloon untenable. Venus is not an easy place to explore. The key to the viability of this design is the separation of hardware elements to operate in environments that do not require leaps in technology. The anchor balloon stays at high altitude, obviating the need to design a balloon (such as the VME) that can survive at a temperature of 350C (700F).

The overall architecture of the Calypso Scout is highlighted in the artist's depiction below. In the actual mission the Balloon would be floating above the clouds and haze, but those are eliminated in this illustration for simplicity. In fact, the whole purpose of Calypso is to allow cameras to venture below those clouds and get a clear view of the surface.

This shows the Descent Module ("Bathysphere") on its tether, skimming over mountainous terrain. In reality, the tether will be deployed to a length of 20-40 kilometers. The Deployment Gondola is also shown, suspended a few meters below the balloon. Solar panels are located along the upper rim of the Gondola. At the flotation altitude of 50-55 km, sunshine will provide ample power.

The anchor balloon will be traveling with the winds of Venus. The science module will also be carried along at that velocity, allowing it to conduct a transect of the ground below. The temperature at an altitude of 10 kilometers is 380 C (720F). Insulation and high-temperature electronics will allow the science module to stay at that altitude for an extended period of time.

However, the module must rise after a foray of 5-6 hours. After being reeled in to the anchor balloon, the science module will cool to 50 C, followed by another deep dive.

The bathysphere will be well insulated, but its time at depth is limited by the time required for its interior to reach 150C. That is the limit for state-of-the-art electronics. Calypso aims to limit technology development and will use available avionics. Allowance needs to be made to guard against the effects of droplets of sulfuric acid. This, however, is a well understood technology challenge.

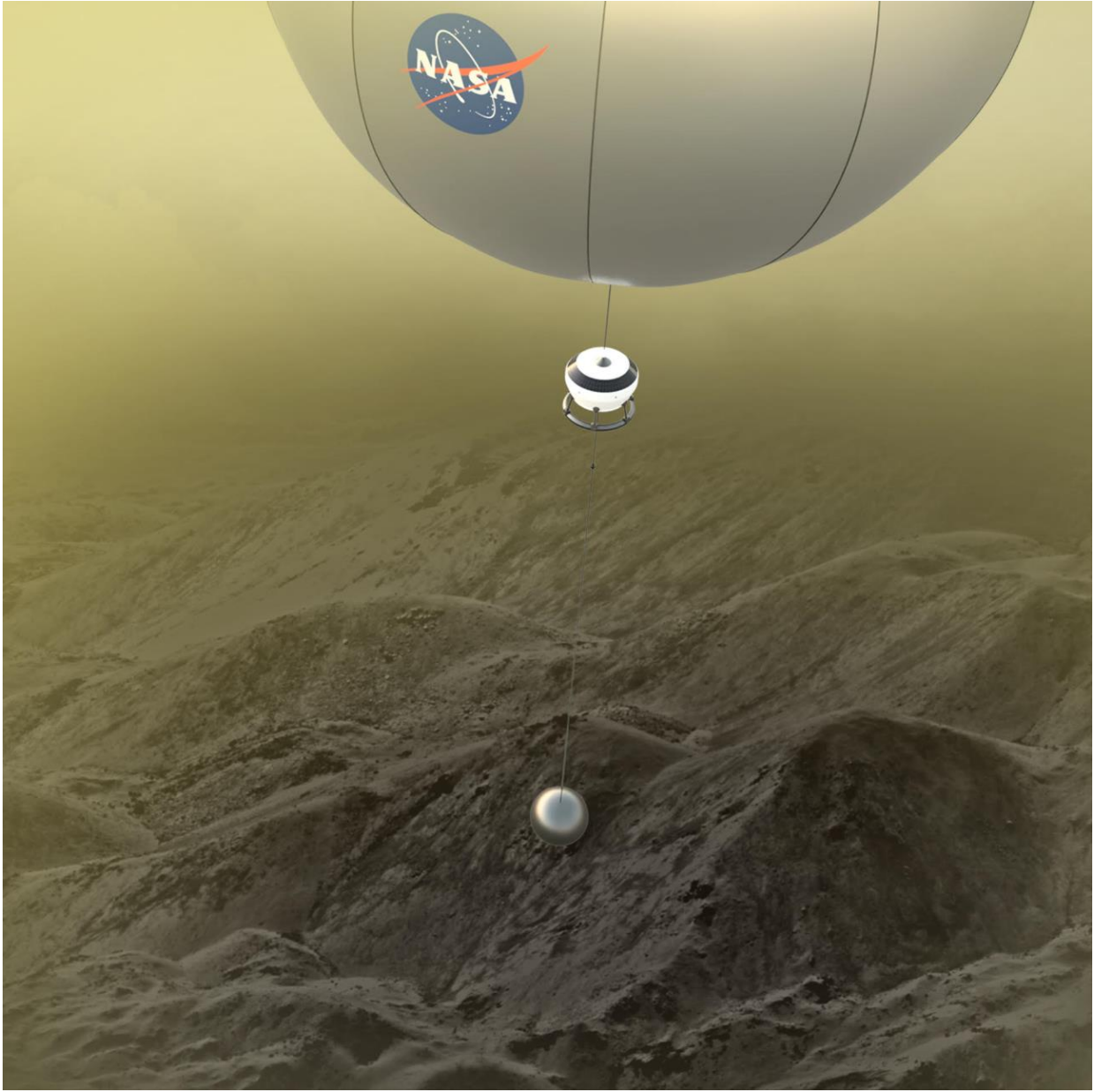

**Calypso Venus Scout**  (credit: Sam Zaref)

**Mission Overview**

  The Calypso Scout will be encapsulated in an Aeroshell during transit to Venus.  Housekeeping, navigation and communications will be handled by an attached Cruise Stage.

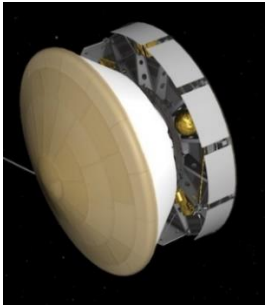
At Venus, the Aeroshell separates from the Cruise Stage and begins its Entry, Descent and Flotation sequence. At an altitude of 75 km, the backshell/parachute combination is ejected, followed by inflation of the Cruise Balloon. Upon descending to 50-55 km, the Balloon will maintain a stable flotation regime. Upon reaching the Flotation stage, the payload in the Descent Module is activated and checked out.

The Descent Module will remain attached to the tether as it is reeled out of the Gondola deployment mechanism. Calypso will demonstrate control of the module during deployment, aerodynamic stability at various altitudes and the ability to collect meaningful science data.

**Payload**

Calypso will carry a High-Resolution Imager and a wide-angle Context Camera. Both are crucial to conducting aerial Field Trips. Below the haze layers, the atmosphere is clear, allowing visible-light cameras to photograph the various terrains found on Venus. The Context Camera will provide an overview of a location, with 1-meter resolution. The narrow-angle Hi-Res camera will allow deeper insight, with images that reach a resolving power of 1-10 cm.

Further insight will be provided by the near-IR imager. It will allow first-order estimates to be made of the mineralogy, and by inference, lithology. The power of Calypso is that these measurements will not be confined to one or two landing sites. Rather, a large number of targets will be surveyed, allowing access to most of Venus' major geological provinces.

The Gondola will include an engineering camera to monitor the operation of the winch. This camera will provide, as a bonus, views of flight within the haze layer. The Gondola could also carry instruments to sample, and analyze, the atmosphere. That bonus science will depend on the funding level.

**Technology**

One of the major expenses of a space mission is the use of new technology. Calypso keeps that to a minimum. It uses legacy hardware and a minimal payload.

**Tether**

The tether is key to the success of the mission. It will connect the Gondola with the Descent Module and will provide deployment support, communications and power. There is a commercial firm that can provide such a tether from existing designs. For Calypso, the tether will undergo extensive testing in simulated T and P environments that it will encounter in the atmosphere of Venus.  This lifeline will need to be 40 kilometers in length.

**Deployment Mechanism**

  The Deployment System consists of a motor-driven storage reel, as well as a winding system for retrieval of the Descent Module.  This hardware will require a thorough program of testing and validation.  The success of the mission depends on this device's ability to deploy and retrieve the bathysphere numerous times.

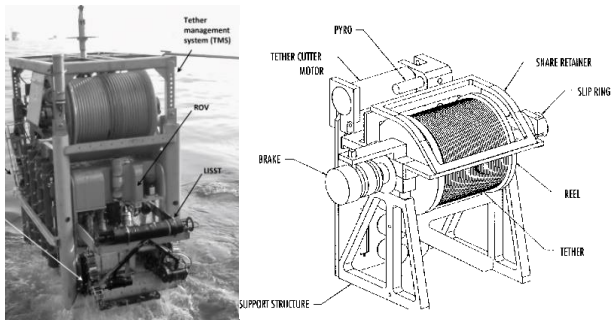

**Examples of Tether Deployment mechanisms**

**Electronics**

Progress in developing high-temperature electronics will be useful in operations in the deep atmosphere of Venus.  Calypso will utilize Hi-T electronics that have a pedigree of use in the industrial world for its instruments.  That limits their use to a temperature of 150C.

  **The Need for High Resolution**

  A variety of models have been proposed to account for the terrain features on Venus.  Plate tectonics, bolide impact and plume upwelling (or downwelling) have been put forth to explain tesserae and/or Ishtar Terra.  The only way to progress forward is with "ground truth."  The Magellan SAR images provide a mid-level view.  The next step is sub-meter observations that can elucidate the detailed stratigraphy of these terranes.  Calypso will provide a series of Landsat quality "aerial road cuts" as it conducts a series of traverses. Here are a few examples of the improved resolution that Calypso will provide.

This SAR image of tesserae on Venus illustrates the resolution from the Magellan mission.

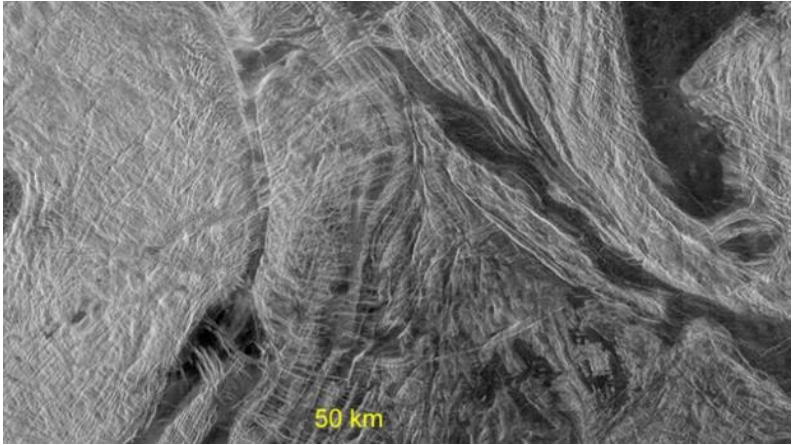

With Calypso, details begin to emerge as the imaging payload descends to lower altitudes. These views of the Hindu-Kush and Canteen Creek Anticline (Australia) illustrate that.

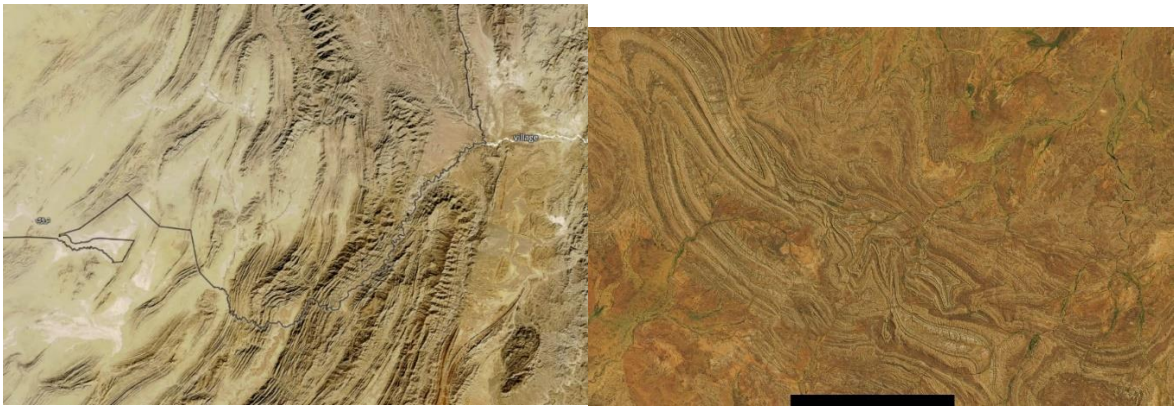

At an altitude of 10 km, the view of the surface will resemble that of Comb Ridge.

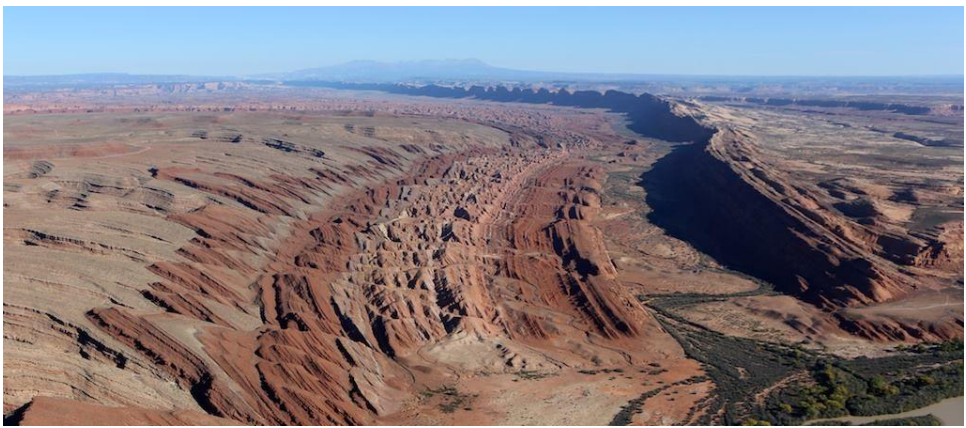

This image of a lava flow on Mars is a preview of Calypso's images as it enters the clear atmosphere at an altitude of 15-25 km.

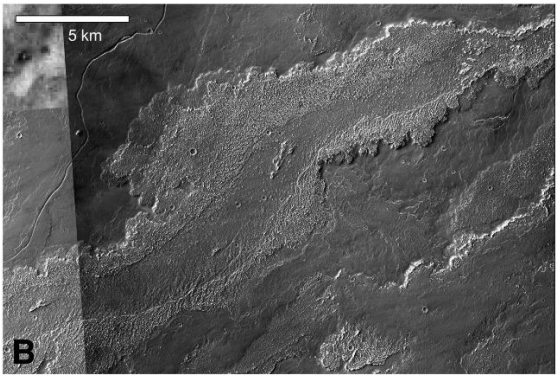

As it approaches a height of 10 km, the view from Calypso will get even better as show by these features on the Earth. Resolutions on the order of centimeters will allow the delineation of individual flows, as well as surface morphology of those flows.

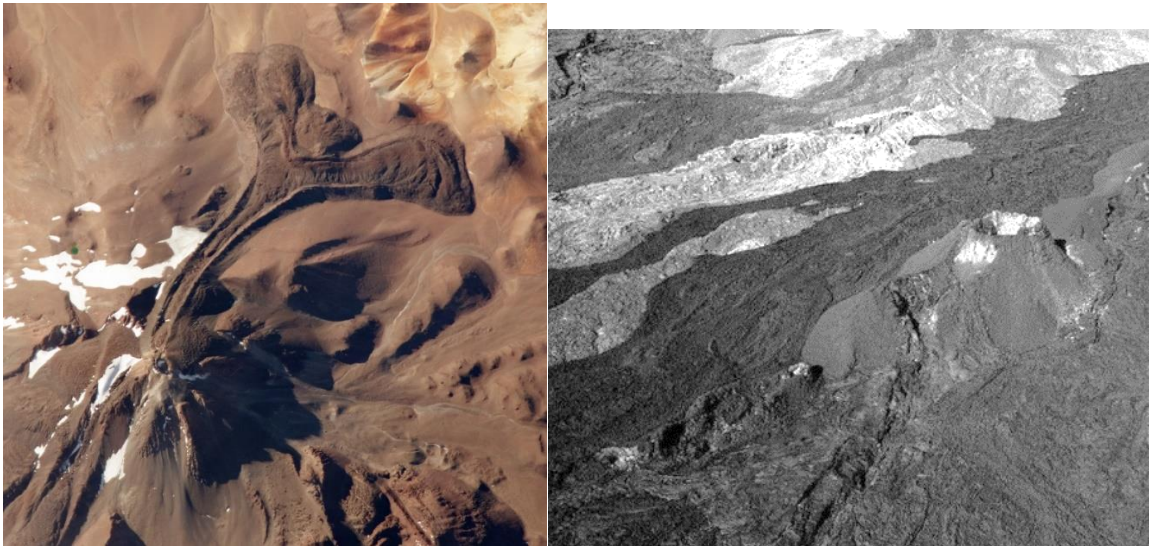

Llullaillaco Volcano, Chile and Mauna Loa

This SAR image of Artemis Chasma is contrasted with a view of the East Africa Rift zone.

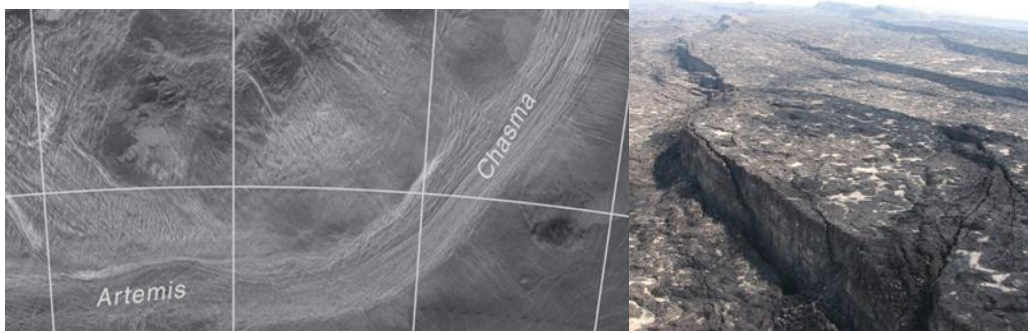

**Future Missions**

   Block II Calypso vehicles will be able to actively steer, or hover, above a site of interest. Block III vehicles will have the ability to set the Instrument Module on the surface for several minutes. This "touch-and-go" operation will allow the collection of samples that can be analyzed at high altitude.  During this brief visit, quick analyses of rocks at the site can be conducted with a Laser-Induced Breakdown Spectrometer (LIBS).  Tests have demonstrated that a Venus-specific LIBS instrument will function on the surface.  This design also provides a pathway for a plausible Venus Sample Return mission.  Soil and rocks can be taken to a waiting Earth-return rocket attached to the high-altitude balloon.  There is no need to launch the vehicle from the surface.

**Conclusions**

   The task of exploring Venus is daunting.   Our sister planet is a paradise for those who study rocks, though its surface resembles the hotter regions of mythology.  The key advantages of Calypso are its mobility and clear views of the surface.  It will provide access to a variety of provinces on the planet, enabling detailed studies of the geology of Venus.   Just as the Bathysphere revealed the depths of Earth's oceans, so Calypso will dive into the depths of Venus atmosphere.  It is not a lander, but it is the next best thing.